\documentclass[twocolumn,amsmath,amssymb]{revtex4}
\usepackage[english]{babel}
\usepackage{graphicx}

\begin{document}

\title{Borrmann Effect in Photonic Crystals: Nonlinear Optical Consequences}

\author{I. Razdolski, T. Murzina, O. Aktsipetrov}

\affiliation{Physics Department, Moscow State University, 119991 Moscow, Russia}

\author{M. Inoue}

\affiliation{Toyohashi University of Technology, 441-8580, Toyohashi, Japan\\}

\begin{abstract}
Nonlinear-optical manifestations of the Borrmann effect that are consequences of the spectral dependence of the spatial distributions of the electromagnetic field in a structure are observed in one-dimensional photonic
crystals. The spectrum of the light self-focusing effect corresponding to the propagation-matrix calculations has been measured near the edge of the photonic gap.
\end{abstract}

\maketitle

Photonic crystals, i.e., microstructures with the periodic
modulation of the refractive index that have a photonic
gap, are actively studied \cite{yablonovitch}, \cite{1dmpmc}. Various nonlinear
optical effects such as the generation of the second- \cite{pellegrini}, \cite{shgmpc} and third-optical harmonics
\cite{eleven} and self-action (selffocusing)
radiation effects \cite{mein} were observed in photonic
crystals.

The Borrmann effect is the anomalous transmission
of x-rays due to the spectral dependence of the spatial
distribution of the electromagnetic field in a crystal.
This effect was observed in quartz \cite{borrmanneins}; more recently, a
similar phenomenon was observed in calcite crystals \cite{borrmannzwei}. The anomalous transmission was explained by von
Laue \cite{laue}: since a crystal body is a periodic atomic structure,
the eigenmodes of the electromagnetic field in an
x-ray range are standing waves. For various wavelengths,
the antinode of a standing wave can be either
on an atom or between the atoms. In the latter case, the
absorption of light in the substance is much lower and,
correspondingly, the transmittance of x-ray radiation is
anomalously high.

It can be assumed that the Borrmann effect should
be observed in photonic crystals. According to the optical
analog of the Bloch theorem, the solution of the
wave equation in a structure with a periodically varying
refractive index is a plane wave modulated in amplitude
with a period coinciding with the modulation period of
the refractive index. Thus, the spectral dependence of
the optical field in a one-dimensional photonic crystal
in the direction perpendicular to the layer plane has
nodes and antinodes whose mutual arrangement certainly
depends on the wavelength of the incident radiation.
By varying the wavelength (or the angle of radiation
incidence for one-dimensional photonic crystals),
one can shift the antinodes of the standing wave from
optically denser layers to optically less dense layers and
observe modifications of various nonlinear optical
effects. In particular, if the nonlinear optical susceptibilities
of the photonic crystal layer are significantly
different, the magnitude of the nonlinear optical effect
depends on the type of layers in which the antinodes of
the standing light wave appear inside the photonic
crystal.

The self-action of light is a process based on change
in the refractive index of the substance under the action
of an intense light field. The radiation self-focusing
effects are quite well studied for the propagation of
intense Gaussian light beams \cite{stryland}. The electromagnetically
induced addition to the refractive index in the
self-focusing effect is quadratic in the field,
$\Delta n \backsim \chi^{(3)}_{(\omega=\omega+\omega-\omega)}|E_{\omega}|^{2}$, where $\chi^{(3)}$ is the third-order nonlinear
susceptibility tensor and $E_{\omega}$ is the electric field
strength of the probe radiation. This effect was taken
for the visualization of the Borrmann effect, because it
is insensitive to the phase of the interacting waves in
contrast to, e.g., the generation of harmonics. Thus, the
Borrmann effect in a nonlinear photonic crystal formed
by alternating linear and nonlinear layers should be
manifested in the spectral dependence of the self-focusing
effect magnitude.

The aim of this work is to observe an optical analog
of the Borrmann effect in nonlinear photonic crystals
using the nonlinear self-action method. The photonic
crystal samples were produced by magnetron sputtering.
A photonic-crystal mirror that is a one-dimensional
photonic crystal consisting of six bilayers — 96-nm bismuth-
doped yttrium iron garnet (Bi:YIG) and 149-nm
silicon oxide SiO$_{2}$ - is deposited on a fused-silica
substrate. The optical thickness of each component of a
bilayer is $\lambda/4$, where $\lambda\simeq 870$ nm is the central wavelength of the photonic gap. The third-order nonlinearity
of yttrium iron garnet in the optical and near IR ranges
is much higher than that of the silicon oxide layers \cite{nonlinearity}. The measured transmission spectrum of the structure is
shown in the inset in Fig. 1a.

The distribution of the electric field in the above
structure was calculated using the propagation-matrix
method \cite{method}. 
The typical distributions of the electric
field squared $|E_{\omega}|^2$ inside the structure at the radiation
wavelength near the spectral edges of the photonic gap
are shown in Fig. 1b, where the spatial distribution of $|E_{\omega}|^2$ in the structure is shown, because the light selffocusing
effect investigated in the experiment is proportional
to this quantity.

 \begin{figure}
  \begin{centering}
  \includegraphics[width=\columnwidth]{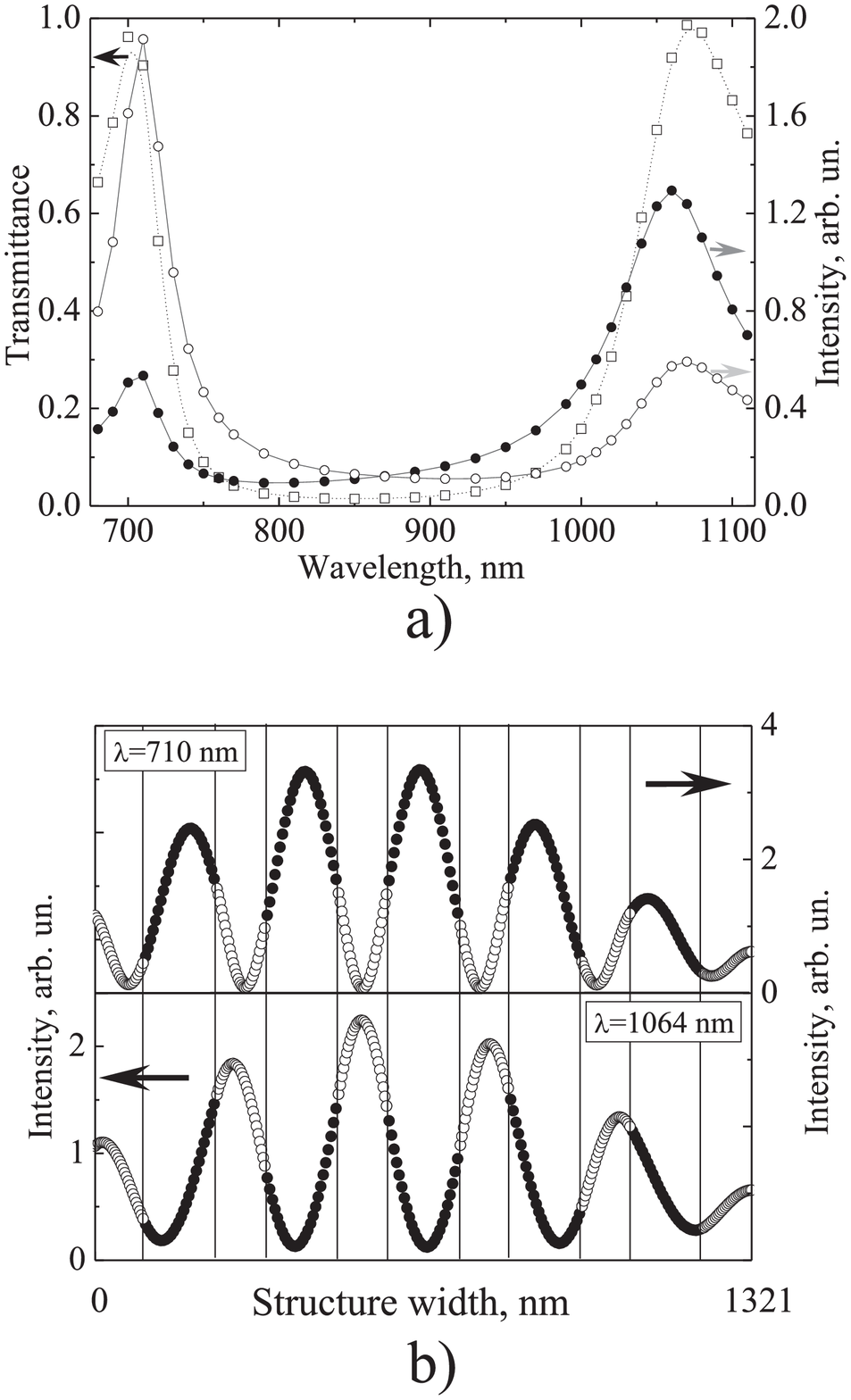}
   \caption{(a) (Open squares, dotted line) Transmittance spectrum
of the photonic crystal and the calculated intensity of
the optical field localized in (open circles) linear and (closed
circles) nonlinear layers of the photonic crystal. (b) The
spatial distribution of the optical field squared in the (closed
circles) linear and (open circles) nonlinear layers of the photonic
crystal for a wavelength of (upper panel) 710 and
(lower panel) 1064 nm as calculated using the propagation
matrix method.}
  \end{centering}
\end{figure}

The radiation localization degree in layers of each
type is determined by summing $|E_{\omega}|^2$ over 30 points in
each layer. The distributions of $|E_{\omega}^{L}|^2$ and $|E_{\omega}^{NL}|^2$, where $|E_{\omega}^{L}|^2$ and $|E_{\omega}^{NL}|^2$ characterize the radiation intensities in the linear (SiO$_{2}$) and nonlinear (Bi:YIG) layers of a
photonic crystal, respectively, were obtained in this
way. Figure 1a shows the spectral dependences of the
radiation intensities $|E_{\omega}^{L}|^2$ and $|E_{\omega}^{NL}|^2$ in the wavelength range covering the photonic gap and its both edges. It is
seen that light at the (upper panel, Fig. 1b) short- and
(lower panel, Fig. 1b) long-wavelength edges of the
photonic gap is predominantly localized in the linear
and nonlinear layers of the photonic crystal, respectively.

The relative radiation intensity $|E_{\omega}^{NL}|^2$ is nonmonotonic
and has a maximum near the long-wavelength
edge. The intensity $|E_{\omega}^{L}|^2$ has a maximum near the
short-wavelength edge. The intensity in the center of
the photonic gap is equally distributed between the linear
and nonlinear layers. The spectral vicinity of the
long-wavelength edge of the photonic gap is nonmonotonic
and has a maximum.

The z-scan method was used to analyze the
cubic self-action effects at the probe radiation wavelength.
The method proposed in \cite{sheik-bahae} is similar to that
used in \cite{mein}. Radiation from a
Nd$^{3+}$ laser (a wavelength
of 1064 nm, a pulse duration of 15 ns, a repetition rate
of 25 Hz, and a pulsed power density up to 10
MW/cm$^{2}$) was focused on a sample by a lens with a focal length
of 6 cm. By means of a translator, the sample is displaced
along the beam-propagation direction near the
focal plane of the lens, allowing one to control the radiation
power density on the sample.

When the self-focusing effect is investigated, radiation
passed through the photonic crystal fell on a limiting
diaphragm and an IKS-1 filter and was detected by
a UPD-70-IR2 photodiode. The effective transmittance
of the aperture receiver was measured in the experiment
as a function of the sample position with respect to the
lens focus, $T(z)$, 
where $T(0)$ is the transmittance of the
sample located near the focal plane of the lens and $T=1$
is the $T$ value for the sample located far from the lens
focus. The nonlinear (two-photon) absorption was measured
similarly, but the diaphragm in front of the photodiode
was in the open position.

 \begin{figure}
  \begin{centering}
  \includegraphics[width=\columnwidth]{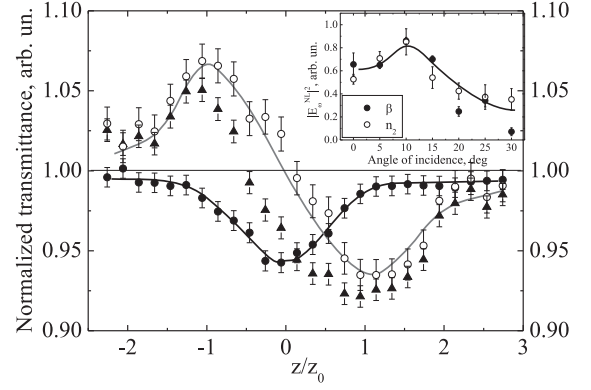}
   \caption{Effective transmittance for the cases of the (closed
circles) open diaphragm, (closed triangles) closed diaphragm,
and (open circles) their ratio. The inset shows the
angular dependence of the intensity of the radiation localized
in the nonlinear layers of the photonic crystal, $|E_{\omega}^{NL}|^2$. The open and closed circles are determined from the selffocusing
and two-photon absorption data, respectively.}
  \end{centering}
\end{figure}

Figure 2 shows the effective transmittance for
z-scan in the cases of the open and closed diaphragms.
To take into account the influence of
absorption on the self-action effect in the case of the
closed-aperture z-scan, the effective transmittance was
normalized over the absorption curve \cite{sheik-bahae}. The normalization
result having a typical shape of the closed-aperture z-scan is also shown in Fig. 2. Similar measurements were carried out at the angles of incidence
from 0° to 30°, corresponding to wavelengths from 1064 to 1090 nm. The typical $n_2$ values ($\Delta n=n_{2}I$, where I is the radiation intensity) in this spectral range were $(3 \div 7)\cdot 10^{-8}$ cm$^2$/W, which are higher than a similar value for magnetophotonic microcavities with the: (Ti$_{2}$O$_{5}$/SiO$_{2}$)$^{5}$/Bi:YIG/(Ti$_{2}$O$_{5}$/SiO$_{2}$)$^{5}$ structure \cite{mein}.
Assuming that the nonlinear susceptibility $\chi^{(3)}_{(\omega=\omega+\omega-\omega)}$ and two-photon absorption coefficient $\beta$ are constant in this spectral range, we determine the relative $|E_{\omega}^{NL}|^2$ values from the $T(z)$ dependencies
obtained for the self-focusing and two-photon absorption
effects in this spectral range.

The angular spectrum of the nonlinear optical
effects (proportional to the pump-field intensity $I_{\omega}^{NL}$) is
recalculated to the frequency spectrum by means of the
formula
$\lambda=\lambda_0(1-n^{-2}_{YIG}sin^{2}\theta)^{1/2}$. The magnitude of
these effects depends on the spatial distribution of the
radiation intensity inside the structure and on the transmittance
of the photonic crystal: $\Delta n=n_{2}I^{NL}\propto|b(\lambda)|^{2}\cdot T(\lambda)$, where $b(\lambda)$ is the Borrmann coefficient
characterizing the degree of the electric field localization
in the nonlinear layers of the structure: $E_{\omega}^{NL}=b(\lambda) t(\lambda)E_{\omega}^{in}$, where $E_{\omega}^{in}$ is the pump field at the input of the photonic crystal and, $t(\lambda)$ is the transmittance for the
electric field, $|t(z)|^2=T(z)$. Since the transmittance in
the region available for measurement, i.e., at the edge of
the photonic gap, varies very abruptly, the curves were
normalized to the transmittance of the photonic crystal $T(\lambda)$, presented in Fig. 1. The spectral dependence obtained for the square of the absolute value of the Borrmann coefficient, $|b(\lambda)|^2$, which is proportional to $|E_{\omega}^{NL}|^2$, is presented in Fig. 3, where the qualitative
agreement between the calculations and measured
quantities is observed.

 \begin{figure}
  \begin{centering}
  \includegraphics[width=\columnwidth]{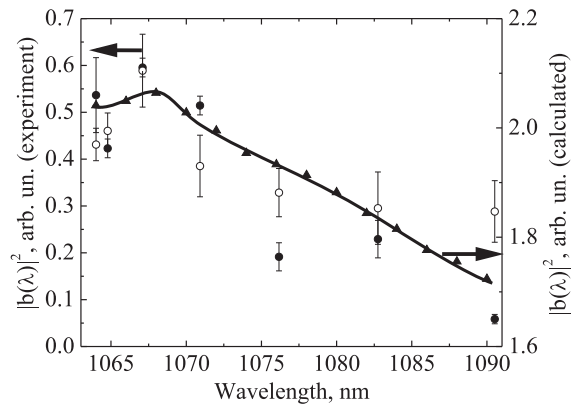}
   \caption{Spectral dependence of the Borrmann coefficient $|b(\lambda)|^2$. The closed and open points correspond to the twophoton
absorption and self-focusing radiation experiments,
respectively. The triangles show the calculated spectral
dependence of $|b(\lambda)|^2$.}
  \end{centering}
\end{figure}

The effect of the spectral dependence of the transmittance
on $|b(\lambda)|^2$ can be excluded by comparing the
self-focusing and two-photon absorption values at the
same transmittances of the photonic crystal, e.g., at different
edges of the photonic gap. Since the edges of the
photonic gap correspond to the localization of the antinodes
of the standing light wave in the photonic-crystal
layers of different types, the self-focusing effect at one
edge (in this case, at the right edge, as seen in Fig. 1a)
is much higher than a similar effect at the opposite edge
of the photonic gap. In this case, the difference between
the localizations of the light field in the linear and nonlinear
layers of the photonic crystal at different edges of
the photonic gap should be most pronounced. At the
same time, the spectral dependencies of the real and
imaginary parts of the refractive indices of the constituent
substances of the photonic crystal should strongly
affect the observed effects.

In summary, the self-focusing and two-photon
absorption effects have been observed in nonlinear photonic
crystals. The spectral dependence of the Borrmann
coefficient, which describes the distribution of
the radiation electric field in the structure, is revealed in these crystals. The experimental data qualitatively coincide
with the calculated dependencies and exhibit the
nonmonotonic spectral dependence of the Borrmann
coefficient near the long-wavelength edge of the photonic
gap. 

The further prospects of the investigations of the
optical Borrmann effect are associated with a comparison
of the nonlinear effects at different edges of the photonic gap for the same transmittance of the photonic crystal.

This work was supported by the Russian Foundation
for Basic Research (Grants \#\# 07-02-91352, 07-02-01358, 06-02-91201).

\vfill\eject




\end{document}